\documentclass[aps,twocolumn,amsmath,amssymb,prl,superscriptaddress,preprintnumbers,nofootinbib,10pt]{revtex4-1}
\usepackage{graphicx,subfigure}
\usepackage{pictexwd,dcpic}

\def\del {\partial}

\begin{document}
\title{Six-Dimensional Superconformal Theories and their Compactifications\\ from Type IIA Supergravity}
\author{Fabio Apruzzi}
\affiliation{Institut f\"ur Theoretische Physik, Leibniz Universit\"at Hannover, Appelstra\ss e 2, D-30167 Hannover, Germany}
\author{Marco Fazzi}
\affiliation{Physique Th\'eorique et Math\'ematique, Universit\'e Libre de Bruxelles, Campus Plaine C.P.~231, B-1050 Bruxelles, Belgium and International Solvay Institutes, B-1050 Bruxelles, Belgium }
\author{Achilleas Passias}
\author{Andrea Rota}
\author{Alessandro Tomasiello}
\affiliation{Dipartimento di Fisica, Universit\`a di Milano--Bicocca, Piazza della Scienza 3, I-20126 Milano, Italy and \\ INFN, sezione di Milano--Bicocca, I-20126 Milano, Italy}

\begin{abstract}
\noindent We describe three analytic classes of infinitely many AdS$_d$ supersymmetric solutions of massive IIA supergravity, for $d=7,5,4$. The three classes are related by simple universal maps. For example, the AdS$_7\times M_3$ solutions (where $M_3$ is topologically $S^3$) are mapped to AdS$_5\times \Sigma_2\times M_3'$, where $\Sigma_2$ is a Riemann surface of genus $g\ge 2$ and the metric on $M_3'$ is obtained by distorting $M_3$ in a certain way. The solutions can have localized D6 or O6 sources, as well as an arbitrary number of D8-branes. The AdS$_7$ case (previously known only numerically) is conjecturally dual to an NS5-D6-D8 system. The field theories in three and four dimensions are not known, but their number of degrees of freedom can be computed in the supergravity approximation. The AdS$_4$ solutions have numerical ``attractor'' generalizations that might be useful for flux compactification purposes. 
\end{abstract}
\maketitle

It is challenging to define renormalizable quantum field theories in high dimensions. The usual Yang-Mills interaction is irrelevant in $d>4$,  as Einstein-Hilbert in $d>2$. Scalar interactions are irrelevant in $d>6$. Some higher-dimensional field theories are known to exist thanks to indirect arguments, e.g., from string theory. The most famous example is the six-dimensional conformal theory describing the dynamics of $N$ coincident M5-branes. Even though its Lagrangian is not known, this theory is expected to enjoy $(2,0)$ supersymmetry, and to have $\sim N^3$ degrees of freedom. 

Understanding higher-dimensional field theories is thus an interesting problem for both quantum field theory and string theory. Moreover, one can expect interesting phenomena upon compactification to lower dimensions. For example, compactifying the aforementioned $(2,0)$ theory on a Riemann surface $\Sigma_g$, one breaks conformal invariance, but the resulting four-dimensional theory flows in the infrared to an ${\cal N}=2$ superconformal field theory (SCFT). Theories obtained this way have interesting duality properties encoded by $\Sigma_g$ \cite{gaiotto,witten-D4}. 

In this Letter, we will study six-dimensional SCFTs with $(1,0)$ supersymmetry. As we will see, these theories exhibit many of the features of their more symmetric (and thus more constrained) $(2,0)$ cousin. For example, they have cubically scaling degrees of freedom, and they give rise upon compactification to interesting SCFTs in four (and three) dimensions.

We will study these models using the AdS/CFT correspondence. The $(2,0)$ theory is dual to the solution AdS$_7\times S^4$ of M-theory; we will look instead at AdS$_7$ solutions in type II theories. These were classified recently \cite{afrt}. If the Romans mass parameter $F_0$ vanishes, the only solutions are reductions of orbifolds of AdS$_7\times S^4$ by ${\Bbb Z}_k$ or $D_k$. With $F_0\neq 0$, infinitely many new solutions were obtained numerically. These were later argued \cite{gaiotto-t-6d} to be near-horizon limits of NS5-D6-D8-brane intersections, considered long ago in \cite{hanany-zaffaroni-6d,brunner-karch}. 

The holographic duals of the compactifications of the $(2,0)$ theory can be obtained by replacing
 AdS$_7$ with either AdS$_5\times \Sigma_2$ \cite{maldacena-nunez} or AdS$_4\times \Sigma_3$ \cite{pernici-sezgin,acharya-gauntlett-kim,gauntlett-macconamnha-mateos-waldram}, where $\Sigma_2$ is a Riemann surface and $\Sigma_3$ is a maximally symmetric space. In both these cases the internal $S^4$ is also distorted in a certain way.

It is natural to wonder whether a similar process can also be applied to the AdS$_7$ solutions of \cite{afrt}. This would indicate that the corresponding $(1,0)$ SCFTs give rise upon compactifications to SCFTs in four and three dimensions, just like for the $(2,0)$ theory. In recent work \cite{afpt,rota-t} we found that this can indeed be done. In the process of doing so, we were able to find analytic expressions for the AdS$_7$ solutions of \cite{afrt} themselves, and analytic maps $\psi_5$, $\psi_4$ from those to the AdS$_5\times \Sigma_2$ and AdS$_4\times \Sigma_3$ solutions. These maps are invertible and they can, of course, be composed: 
\begin{equation}
\begin{split}
\begindc{\commdiag}[450]
\obj(1,1)[a]{${\rm AdS}_7 $} 
\obj(0,0)[b]{${\rm AdS}_4\times\Sigma_3$} 
\obj(2,0)[c]{${\rm AdS}_5\times\Sigma_2$\ .} 
\mor{a}{b}{}[+1,11]
\mor{a}{c}{}[+1,11]
\mor{b}{c}{}[-1,11]
\enddc
\end{split}
\end{equation}
So in the end we have three new classes of infinitely many new backgrounds with analytic expressions, holographically dual to SCFTs in six, four, and three dimensions, with respectively eight, four, and two $Q$-supercharges. The AdS$_4$ vacua also have potential applications to flux compactifications. 

The AdS$_7$ solutions have the following general form. The internal space $M_3$ is an $S^2$ fibration over an interval, parametrized by a coordinate $r \in [r_-,r_+]$:
\begin{equation}\label{eq:ads7}
	e^{2A} ds^2_{{\rm AdS}_7} + dr^2 + e^{2A} v^2 ds^2_{S^2}\ .
\end{equation}
$A$ (the ``warping'') and $v$ are functions of $r$; so is the dilaton $\phi$. We will see below how these three functions are fixed by the equations of motion and preserved supersymmetry. The $S^2$ has a round metric, and its isometry group is the SU(2) R-symmetry of the $(1,0)$ SCFT$_6$. It shrinks at the endpoints $r_\pm$ of the interval. The fluxes have all the components compatible with the R-symmetry: $F_0$, $F_2\propto {\rm vol}_{S^2}$, $H \propto dr \wedge {\rm vol}_{S^2}$. 

The map $\psi_4$ takes the metric, Eq.\ (\ref{eq:ads7}), to 
\begin{equation}\label{eq:74}
	\sqrt{\frac{5}{8}} \left[ \frac58 e^{2A} \left( ds^2_{\rm AdS_4} + \frac{4}{5} ds^2_{\Sigma_3} \right) + dr^2 + \frac{e^{2A} v^2}{1-6v^2} Ds^2_{S^2}\right]
\end{equation}
with $\Sigma_3$ a compact quotient of hyperbolic space, normalized so that its scalar curvature is $-6$. $S^2$ is now fibered over $\Sigma_3$, in a way associated to its tangent bundle; even though the $S^2$ is still round, the total internal space has no isometries. The solution has now four supercharges; it is dual to an ${\cal N}=1$ SCFT$_3$. The fluxes now acquire also components along $\Sigma_3$. The dilaton $\phi_7$ of the AdS$_7$ solutions is taken to $\phi_4$ given by
\begin{equation}
	e^{\phi_4} = \left(\frac58\right)^{1/4} \frac{e^{\phi_7}}{\sqrt{1-6v^2}} \ .
\end{equation}

Similarly, the map $\psi_5$ takes the metric, Eq.\ (\ref{eq:ads7}), to 
\begin{equation}\label{eq:75}
	\sqrt{\frac34} \left[ \frac34 e^{2A} \left( ds^2_{\rm AdS_5} + ds^2_{\Sigma_2} \right) + dr^2 + \frac{e^{2A} v^2}{1-4v^2} Ds^2_{S^2}\right]
\end{equation}
with $\Sigma_2$ a Riemann surface, again normalized so that its scalar curvature is $-6$. $S^2$ is fibered over $\Sigma_2$ via one of its U(1) isometries; in other words, it can be written as ${\Bbb P}({\cal K } \oplus {\cal O})$, and actually ${\cal K}$ is the canonical bundle of $\Sigma_2$. The isometry group is now this U(1), which is the R-symmetry of the ${\cal N}=1$ superalgebra of the SCFT$_4$. Again the fluxes acquire components along $\Sigma_2$. The dilaton $\phi_7$ of the AdS$_7$ solutions is taken to $\phi_5$ given by
\begin{equation}
	e^{\phi_5} = \left(\frac34\right)^{1/4} \frac{e^{\phi_7}}{\sqrt{1-4v^2}} \ .
\end{equation}

So far we have described how the AdS$_7$ solutions get mapped to AdS$_4$ and AdS$_5$ solutions. Remarkably, the AdS$_5$ system of equations turned out much simpler than the ones in AdS$_7$ and AdS$_4$; so much so that we were able to integrate it analytically. (This simplicity ultimately results from a more general classification effort in \cite{afpt}, where existence of AdS$_5$ solutions is reduced to a set of partial differential equations; these simplify and become solvable with an inspired Ansatz.) We can then use the maps, Eqs.\ (\ref{eq:74}) and (\ref{eq:75}), to produce AdS$_7$ and AdS$_4$ solutions as well. In what follows, we will describe the AdS$_7$ solutions. 

Let us first give the simplest example. The metric can be written as 
\begin{equation}\label{eq:d6met}
\begin{split}
		\frac{n_{\rm D6}}{F_0}\sqrt{\tilde y+2}&\left(\frac43ds^2_{{\rm AdS}_7}+\right.\\
 &\left.\frac{d\tilde y^2}{4(1-\tilde y)(\tilde y +2)} +\frac13 \frac{(1-\tilde y)(\tilde y+2)}{8 - 4 \tilde y - \tilde y^2} ds^2_{S^2}\right)
\end{split}
\end{equation}
where $\tilde y\in [-2,1]$. The flux can be found in \cite{afpt}; the dilaton is given by $e^{\phi_7}=\frac2{\sqrt{n_{\rm D6}F_0}}\frac{(\tilde y+2)^{3/4}}{\sqrt{8-4\tilde y - \tilde y^2}}$. Around $\tilde y=1$, the internal metric is $\sim\frac{d \tilde y^2}{4(\tilde y-1)}+(\tilde y-1) ds^2_{S^2}$, which turns into flat space $d \rho^2 + \rho^2 ds^2_{S^2}$ by the change of coordinates $\rho= \sqrt{\tilde y-1}$. So $\tilde y=1$ is a regular point. On the other hand, around $\tilde y=-2$ the metric behaves as $\sim 16\sqrt{\rho}ds^2_{{\rm AdS}_7}+ \frac1{\sqrt \rho}(d \rho^2 + \rho^2 ds^2_{S^2})$, with $\rho = \tilde y + 2$, which is the correct behavior near a stack of D6-branes wrapping AdS$_7$. 

\begin{figure}[ht]
\centering	
	\subfigure[\label{fig:d6}]{\includegraphics[scale=.4]{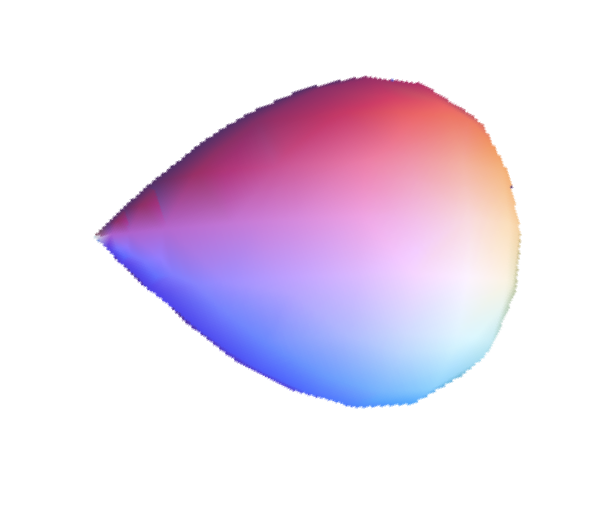}}
	\hspace{.4cm}
	\subfigure[\label{fig:ns5d6}]{\includegraphics[scale=.5]{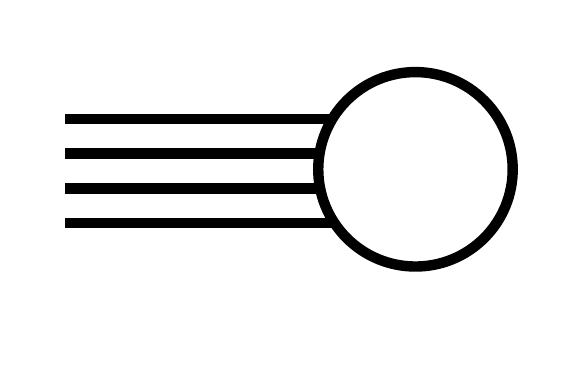}}
	\caption{In \subref{fig:d6}, a sketch of the internal $M_3$ in Eq.\ (\ref{eq:d6met}); the cusp represents the single D6 stack. In \subref{fig:ns5d6}, the brane configuration whose near-horizon should originate, Eq.\ (\ref{eq:d6met}). The dot represents a stack of $N$ NS5-branes; the horizontal lines represent $n_{\rm D6}$ D6-branes ending on them.}
	\label{fig:d6sketch}
\end{figure}

A stack of electric charges of the same sign in a compact space would be in contradiction with Gauss' law, but in type IIA this reads $d F_2 - H F_0 = \delta$. Integrating this, we get $N n_0 = n_{\rm D6}$, where $N=\frac1{4\pi^2}\int H$ and $n_0 = 2\pi F_0$ are integers by flux quantization. The corresponding brane configuration should consist of $n_{\rm D6}$ D6-branes ending on $N$ NS5-branes. Notice that the net number of D6-branes ending on an NS5 is $n_0$; this again gives $N n_0 = n_{\rm D6}$. 

More general solutions can be described as follows. Given a solution $\beta(y)$ to the ordinary differential equation
\begin{equation}\label{eq:ode}
	\del_y(q^2)= \frac29 F_0 \ ,\qquad q\equiv -\frac{4y\sqrt{\beta}}{\del_y \beta}\ ,
\end{equation}
the warping $A$, dilaton $\phi$, and volume function $v$ can be described by 
\begin{equation}\label{eq:Abeta}
\begin{split}
		e^{A}= \frac23 \left(-\frac {\del_y\beta}y\right)^{1/4}&\ ,\qquad e^{\phi}=\frac{(-\del_y\beta/y)^{5/4}}{12\sqrt{4 \beta - y \del_y\beta}}\ ,\\
		v^2=&\frac{\beta/4}{4 \beta-y \del_y\beta}\ .
\end{split}
\end{equation}
Moreover, $dr=\left(\frac34\right)^2 \frac{e^{3A}}{\sqrt \beta} dy$. 

The local behavior of a solution can be read off from $\beta$ as follows: 
\begin{itemize}
	\item At a single zero of $\beta$, the $S^2$ shrinks in a regular way, so that one has a regular point (as was $\tilde y=1$ in Eq.\ (\ref{eq:d6met})).
	\item At a double zero of $\beta$, there is a stack of D6-branes (as at $\tilde y=-2$ in Eq.\ (\ref{eq:d6met})).
	\item At a square root point, where $\beta\sim \beta_0 + \sqrt{y-y_0}$, there is an O6 singularity. 
\end{itemize}

Equation (\ref{eq:ode}) can be easily integrated by writing $16 y^2 \frac{\beta}{(\beta')^2}= \frac29 F_0 (y-\hat y_0)$, for $\hat y_0$ a constant; this can now be integrated by quadrature. For $F_0=0$, the solution can be written as
\begin{equation}\label{eq:beta0}
	\beta= c(y^2-y_0^2)^2\ .
\end{equation}
Using Eq.\ (\ref{eq:Abeta}), this gives rise to the solution discussed in Sec.\ 5.1 of Ref.\ \cite{afrt}, which is obtained by reducing AdS$_7\times S^4$ to IIA. (In that paper $c=\frac4{k^2}$, $y_0=\frac9{32}R^3$.) Equation (\ref{eq:beta0}) has two double zeros; hence, the corresponding solution has two D6 stacks, one at $y=y_0$ with D6-branes and one at $y=-y_0$ with an equal number of anti-D6-branes. These originate from loci where the reduction from M-theory degenerates.

For $F_0\neq 0$, the simplest solution reads
\begin{equation}\label{eq:betanice}
	\beta= \frac8{F_0}(y-y_0)(y+2y_0)^2\ .
\end{equation}
This has a single zero in $y=y_0$ and a double zero in $-2y_0$; so it has only one stack of D6-branes. Flux quantization requires $\int_{S^2} \tilde F_2 = n_{\rm D6}$, where $\tilde F_2 = F_2 - B F_0$ is the closed Ramond-Ramond form; in this case it fixes 
\begin{equation}\label{eq:y0F0}
	y_0 = -\frac 38 \frac{n_{\rm D6}^2}{F_0}\ .
\end{equation}
Substituting Eq.\ (\ref{eq:betanice}) in Eq.\ (\ref{eq:Abeta}), defining $\tilde y \equiv \frac y{y_0}$, and using Eq.\ (\ref{eq:y0F0}), we reproduce Eq.\ (\ref{eq:d6met}), which indeed has only one D6 stack. 

We will now describe more general solutions. The first generalization will introduce either a second D6 stack, or an O6 singularity. The second generalization will involve D8-branes. These were described numerically in \cite{afrt,gaiotto-t-6d}, but we will now be able to give analytic expressions. It would also be possible to combine D6, D8, and O6 into even more general solutions.

The first generalization involves finding a more general $\beta$ that solves Eq.\ (\ref{eq:ode}) for $F_0 \neq 0$. This can be written as 
\begin{equation}\label{eq:gen}
	\beta = \frac{y_0^3}{b_2^3 F_0}\left(\sqrt{\hat y} - 6\right)^2\left(\hat y + 6 \sqrt{\hat y}+6b_2 - 72 \right)^2 \ , 
\end{equation}
where
\begin{equation}
	\hat y \equiv 2 b_2 \left(\frac y{y_0} -1\right)+36\ . 
\end{equation}
The parameter $b_2$ is also equal to $\frac{F_0}{y_0} \beta_2$, where $\beta_2$ is half the second derivative of $\beta$ in $y=y_0$. 
\begin{itemize}
	\item If $b_2<12$, $\beta$ has two double zeros, so the solution corresponds to two D6 stacks, one at $\sqrt{\hat y} = -3+\sqrt{81-6 b_2}$ and one at $\hat y = 36$. 
	\item If $b_2 >12$, the solution corresponds to a D6 stack at one pole $\hat y=36$ and an O6 singularity at $\hat y= 0$.  
	\item If $b_2 = 12$, $\beta$ simplifies to $\frac{y_0^3}{1728 F_0} \hat y (\hat y-36)^2$, which is Eq.\ (\ref{eq:betanice}) up to coordinate change; so this case corresponds to a single D6 stack at $\hat y=36$. 
\end{itemize}

The second generalization consists in introducing D8-branes. These manifest themselves as loci across which $F_0$ (and hence Eq.\ (\ref{eq:ode})) can jump. Supersymmetry requires them to wrap the round $S^2$ in Eq.\ (\ref{eq:ads7}) at a fixed $r=r_{\rm D8}$; this is indeed the only way they can preserve the SU(2) R-symmetry. The supergravity solutions consist in gluing together solutions of the following type: Eqs.\ (\ref{eq:betanice}), (\ref{eq:gen}), or	(\ref{eq:beta0}); the only non-trivial task is fixing the parameters of those solutions, and the positions of the D8-branes, using flux quantization. We will do so for an example with one D8-brane and one example with two D8-branes; here Eq.\ (\ref{eq:gen}) will not be needed, but we expect it to become relevant for higher numbers of D8-branes.

A D8-brane can also have D6-brane charge $\mu$ smeared on its worldvolume; this is the Chern class of a gauge bundle, and as such it is an integer. D8-branes with the same $\mu$ will be stabilized by supersymmetry on top of each other. In such a situation, the flux integers of $F_0$ and $F_2$ before and after the D8-brane stack, $(n_0,n_2)$ and $(n_0',n_2')$, are related to the number of branes in the stack and their charge by $n_{\rm D8}=n_0'-n_0$ and $\mu = \frac{n_2'-n_2}{n_0'-n_0}$. The position is then fixed by the formula \cite{afrt,gaiotto-t-6d}
\begin{equation}\label{eq:qd8}
	q |_{r=r_{\rm D8}}= \frac{n_2' n_0 - n_2 n_0'}{2(n_0'-n_0)} =  \frac12(-n_2 + \mu n_0) = \frac12 (-n_2' + \mu n_0') \ ,  
\end{equation}
where $q$ was given in Eq.\ (\ref{eq:ode}). So we see that in the $y$ coordinate the position of the D8-branes goes quadratically with $\mu$. In fact, $q$ itself has a nice interpretation: from its definition, Eq.\ (\ref{eq:ode}), and from Eqs.\ (\ref{eq:Abeta}) and (\ref{eq:ads7}) we see 
\begin{equation}
	q = \frac14 v e^{A-\phi}= e^{-\phi} {\rm radius}(S^2)\ .
\end{equation}

The simplest possibility is to have one D8 stack, of charge $\mu$. This is done by gluing two copies of Eq.\ (\ref{eq:d6met}). Concretely, the function $\beta$ reads 
\begin{equation}
	\beta= \left\{  
	\begin{aligned}
		\frac8{F_0}(y-y_0)(y+2y_0)^2\ , \quad y_0<y<y_{\rm D8} \ ;\\
		\frac8{F_0'}(y-y_0')(y+2y_0')^2\ , \quad y_{\rm D8}<y<y_0' \ ;
	\end{aligned} 
	\right.
\end{equation}
with $y_0<0$, $y_0'>0$. 
We need to impose flux quantization, Eq.\ (\ref{eq:qd8}), and continuity of $\beta$ and its derivative (which, via Eq.\ (\ref{eq:Abeta}), guarantees continuity of $A$, $\phi$, and of the metric). This leads to
\begin{equation}
	\begin{split}
	&F_0'= F_0 \left(1-\frac N\mu\right) \ ,\qquad y_{\rm D8}= 3 F_0 \pi^2 (N-2\mu)(N-\mu)\ ,\\
	&y_0 = -\frac32 F_0 \pi^2 (N^2-\mu^2) \ ,\quad y_0'=\frac32 F_0 \pi^2 (N-\mu)(2 N- \mu)
	\ .	
	\end{split}
\end{equation}
We see now that $\beta$ has a single zero at both endpoints $y_0$ and $y_0'$. So this solution is regular, except of course for the effect of the D8 backreaction; this causes discontinuities in the first derivatives of $A$, $\phi$ and the metric, as any domain wall in general relativity should do.

\begin{figure}[ht]
\centering	
	\subfigure[\label{fig:d8}]{\includegraphics[scale=.4]{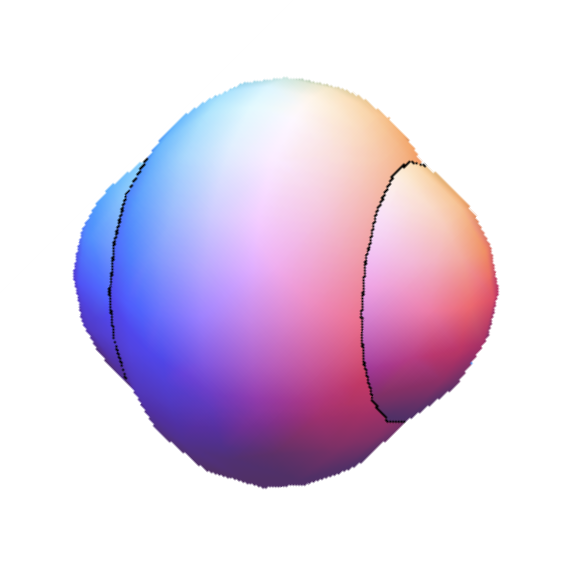}}
	\hspace{.4cm}
	\subfigure[\label{fig:ns5d6d8}]{\includegraphics[scale=.5]{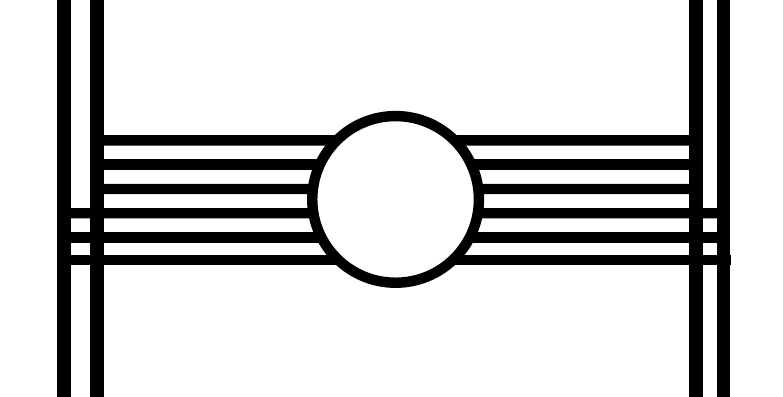}}
	\caption{In \subref{fig:d8}, a sketch of the internal $M_3$ in the solution with two D8-brane stacks, represented by two ``creases''. In \subref{fig:ns5d6d8}, the corresponding brane configuration. The vertical lines represent the D8-branes; each stack has $n_0=2$ branes with $|\mu|=3$.}
	\label{fig:d6d8sketch}
\end{figure}

The next possibility is to have two D8 stacks. As in \cite{afrt,gaiotto-t-6d}, we assume for simplicity that the solution is symmetric under $y\to -y$, so that the two endpoints are at $y_0<0$ and $-y_0$, and the two D8 stacks, of D6 charge $\mu$ and $-\mu$, are located at $y_{\rm D8}<0$ and $-y_{\rm D8}$. There are three regions: (i) For $y_0<y<y_{\rm D8}$,  $F_0>0$; $\beta$ is as in Eq.\ (\ref{eq:betanice}); (ii) for $y_{\rm D8}<y<-y_{\rm D8}$, $F_0=0$, and $\beta$ is as in Eq.\ (\ref{eq:beta0}), namely $\beta=\frac4{k^2}\left[y^2- (\frac 9{32}R^3)^2\right]^2$ and (iii) for $-y_{\rm D8}< y < -y_0$, the Romans mass is $F_0'=-F_0<0$; $\beta$ is again as in Eq.\ (\ref{eq:betanice}), but now with $y_0\to -y_0$, $F_0 \to -F_0$. Again in this way we avoid singularities, except for the discontinuities in the first derivatives induced by the two D8 stacks. This solution, and its brane interpretation, is shown in Fig.\ \ref{fig:d6d8sketch}.

Using flux quantization and Eq.\ (\ref{eq:qd8}) we can fix the parameters as
\begin{equation}\label{eq:2d8q}
	\begin{split}
		y_0 = -\frac94 k &\pi (N-\mu) \ ,\qquad
		y_{\rm D8} = -\frac94 k \pi (N-2 \mu)\ ,\\
		&R^6 = \frac{64}3 k^2\pi^2 (3 N^2 -4\mu^2) \ .\\
	\end{split}
\end{equation}
This solution only exists for $N\ge 2 \mu$, in agreement with a bound in \cite{gaiotto-t-6d}.

All these analytic solutions now allow us to obtain some information about the field theory duals. As we mentioned above, the six-dimensional $(1,0)$ field theories should be dual to the theories described by NS5-D6-D8 configurations such as the ones in Figs. \ref{fig:ns5d6} and \ref{fig:ns5d6d8}. These theories are a bit mysterious because of the physics of coincident NS5-branes. Separating them leads to a quiver description \cite{hanany-zaffaroni-6d,brunner-karch}, but this corresponds to moving along a ``tensor branch'' departing from its origin, the conformal point. Thus, most degrees of freedom are not captured by the quiver description. Using holography, however, we can count them even without a Lagrangian description, also in four and three dimensions. 

A possible measure is given by the coefficient ${\cal F}_{0,d}$ in the free energy as a function of temperature $T$ and volume $V$: ${\cal F}_d= {\cal F}_{0,d} V T^d$. This can be estimated by taking a large black hole in AdS$_{d+1}$: this leads to $\frac{R_{\rm AdS_{d+1}}^5}{G_{{\rm N},d+1}}$, where $G_{{\rm N},d+1}$ is Newton's constant. For constant dilaton this is $\sim \frac{\rm Vol}{g_s^2}$, where ${\rm Vol}$ is the volume of the internal manifold; for us the dilaton is not constant, and we should integrate $e^{-2 \phi}$ over the internal space. All in all, ${\cal F}_{0,6}= \int_{M_3} {\rm vol}_{M_3} e^{5A-2 \phi}$ for the SCFT$_6$, and ${\cal F}_{0,d}= \int_{M_3 \times \Sigma_{6-d}} {\rm vol}_{M_3}\wedge {\rm vol}_{\Sigma_{6-d}} e^{5A-2 \phi}$ for the SCFT$_d$, $d=3,4$. For $d=4$, this is also related to $a$ and $c$ (which are equal up to stringy corrections, which we did not compute). Using the maps, Eqs.\ (\ref{eq:75}) and (\ref{eq:74}), we find the universal relations 
\begin{equation}
	{\cal F}_{0,4}= \left(\frac34\right)^3 \pi  (g-1)  {\cal F}_{0,6}\ ,\quad
	{\cal F}_{0,3}= \left(\frac 58\right)^4  {\rm Vol}(\Sigma_3){\cal F}_{0,6}\ .
\end{equation}

${\cal F}_{0,6}$ can be computed explicitly. As a reference point, in our normalization the $(2,0)$ theory gives ${\cal F}_{0,6}=\frac{128}3 \pi^4 N^3$; for its ${\Bbb Z}_k$ orbifold, this number is multiplied by $k^2$. We have computed ${\cal F}_{0,6}$ for the two cases of Figs.\ \ref{fig:d6sketch} and \ref{fig:d6d8sketch}. For the first case, we simply use the solution, Eq.\ (\ref{eq:d6met}), and we get ${\cal F}_{0,6}= \frac{512}{45} \pi^4 n_{\rm D6}^2 N^3$. For the case of Fig.\ \ref{fig:d6d8sketch}, we have to integrate over the three different regions, recalling Eq.\ (\ref{eq:2d8q}). This gives 
\begin{equation}
	{\cal F}_{0,6}= \frac{128}3 k^2\pi^4 \left(N^3 - 4 N\mu^2 + \frac{16}5 \mu^3\right)\ 
\end{equation}
(in agreement with the approximate result in \cite{gaiotto-t-6d}). For the corresponding SCFT$_4$, 
$a=c=$$\frac{g-1}{3}$$\left(N^3 - 4 N\mu^2 + \frac{16}5 \mu^3\right)$. 

Let us also expand a bit on our earlier comment about the possible uses of the AdS$_4$ solutions for flux compactifications. For many applications (such as to achieve de Sitter vacua, or a hierarchy between the Kaluza-Klein and cosmological constant scales), it is useful to introduce orientifold planes. Yet they are usually artificially ``smeared'', i.e., replaced with a continuous charge distribution. Flux compactifications with localized sources are not many (but see \cite{assel-bachas-estes-gomis}), and the AdS$_4$ solutions that one obtains by applying the map (\ref{eq:74}) to (\ref{eq:gen}) is the first case to our knowledge of solution with a localized O6-plane. Moreover, in Sec.\ 5 of Ref.\ \cite{rota-t} another wide class of AdS$_4$ solutions is uncovered. Although these are so far only known numerically, they appear to be more general; for example, there are solutions with an O6-plane and no further singularity. An attractor mechanism allows us to achieve regularity without need for fine-tuning.

In conclusion, we have found analytic expressions for AdS$_7$ solutions dual to $(1,0)$ SCFTs in six dimensions; we also found AdS$_5$ and AdS$_4$ solutions, related to the AdS$_7$ ones by simple universal maps. This structure strongly suggests that the dual field theories in three and four dimensions are twisted compactifications of the $(1,0)$ SCFT in six dimensions. It would be interesting to further strengthen this case by finding gravity solutions representing the renormalization group flows connecting the AdS$_7$ solution to the lower-dimensional ones. In each of the three classes, one can find infinitely many solutions, each of them characterized by the number and charges of D8- and D6-branes present. The number of degrees of freedom scales cubically with those charges, and with a three-form flux integer playing the role of color. While the $(1,0)$  theories have no Lagrangian description, an effective description is known on a Coulomb branch of their moduli space, where some scalars acquire vacuum expectation values; for more details, see \cite{gaiotto-t-6d,hanany-zaffaroni-6d,brunner-karch}.

\medskip\noindent {\bf Acknowledgements} We would like to thank C.~Bachas, I.~Bah, I.~Bena, N.~Bobev, J.~Maldacena, T.~Van Riet, and A.~Zaffaroni for interesting discussions. F.A.~is grateful to the Graduiertenkolleg GRK 1463 ``Analysis, Geometry and String Theory'' for support. M.F.~is a Research Fellow of the Belgian FNRS--FRS; his work was partially supported by the ERC Advanced Grant ``SyDuGraM'', by IISN-Belgium (convention 4.4514.08) and by the ``Communaut\'e Fran\c{c}aise de Belgique" through the ARC program. A.P., A.R., and A.T.~are supported in part by INFN. A.P.~and A.T. are also supported by the ERC Starting Grant n. 307286 (XD-STRING). The research of A.T.~is also supported by the MIUR-FIRB grant RBFR10QS5J ``String Theory and Fundamental Interactions''.

\bibliography{at}

\end{document}